\documentstyle[11pt]{article}

\def\thebibliography#1{\list
 {$^{\arabic{enumi}}$}{\settowidth\labelwidth{[#1]}\leftmargin\labelwidth
 \advance\leftmargin\labelsep
 \usecounter{enumi}}
 \def\newblock{\hskip .11em plus .33em minus .07em}
 \sloppy\clubpenalty4000\widowpenalty4000
 \sfcode`\.=1000\relax}

\begin{document}

\makeatletter
\renewcommand{\@cite}[2]{{\footnotesize $^{#1\if@tempswa , #2\fi}$}}
\def\ncite#1{{\def\@cite##1##2{##1}\cite{#1}}}
\makeatother

\baselineskip=.7cm

{\large \noindent {\bf THE SEARCH FOR THE ORIGINS OF M-THEORY : } 

 \vspace{3mm}\noindent {\bf LOOP QUANTUM MECHANICS,   LOOPS/STRINGS      }  

 \vspace{3mm}\noindent {\bf AND BULK/BOUNDARY DUALITIES} }

\vspace{5mm}
\begin{quotation}

Carlos Castro$^{*}$
\footnote{Electronic mail address: castro@hubble.cau.edu}

\noindent $^{*}$ {\it Center for Theoretical Studies of Physical Systems,
Clark Atlanta University, Atlanta, Georgia 30314 USA}

\vspace{5mm}
\noindent 
\end{quotation}

\hspace{10mm}

\centerline{\bf ABSTRACT}

The construction of a $covariant$ Loop Wave functional equation in a $4D$ spacetime is attained by introducing a generalized $eleven$ dimensional categorical {\bf C}-space
comprised of $8\times 8$ antisymmetric matrices. The latter matrices encode the generalized coordinates of the histories of points, loops and surfaces $combined$. Spacetime Topology change
and  the Holographic principle are natural consequences of imposing the principle of 
$covariance$ in {\bf C}-space. 
The Planck length is introduced  as a necessary rescaling parameter to establish the correspondence limit with the  physics of point-histories  in ordinary Minkowski space,  in the limit $l_P\rightarrow 0$. 
Spacetime quantization should appear in discrete units of Planck length, area, volume ,....All this seems to suggest that the  generalized principle of covariance,  representing invariance of proper $area$ intervals in {\bf C}-space, under matrix-coordinate transformations, 
could be relevant in discovering the underlying principle behind the origins of $M$ theory. 
We construct an ansatz for  the $SU(\infty)$ Yang-Mills 
vacuum wavefunctional  as a solution of  the Schroedinger Loop Wave equation
associated with  the Loop Quantum Mechanical  formulation of the Eguchi-Schild String . The Strings/Loops ( $SU(\infty)$ gauge field) correspondence   
implements one form of the Bulk/Boundary duality conjecture in this case.

\makeatletter
\renewcommand{\@oddhead}
{{\footnotesize
\hfill
Castro  : M-Theory, Loop Quantum Mechanics, Loops/Strings and  Bulk-Boundary Dualities

 \hfill }}
\makeatother

\newpage

\centerline{\bf 1. INTRODUCTION}   

Recently, a very interesting relation between string quantization based on the Schild string path integral and the Nambu-Goto string path integral was established at the semiclassical level by a saddle point evaluation method \cite{Euro2}. Quantum 
mechanics of Matrices , M(atrix) models  \cite{Ban}theory have been conjectured 
as a leading  
candidate to understand nonperturbative string theory : $M$ theory. 

We will follow another  approach related to ordinary Matrix Quantum Mechanics and 
study Quantum Mechanics but in Loop spaces. The Eguchi quantization of the Schild string is essentially a sort of quantum mechanics formulated in the space of loops. The authors \cite{Euro2} have constructed a Loop Wave equation associated with the Eguchi-Schild string quantization in the Schroedinger form, where the $area$ spanned by the evolution of a closed spatial string ( loop) served as the role of the ``time parameter'' and the holographic shadows of the loop-shapes,  onto the spacetime coordinate planes, $\sigma^{\mu\nu}(C)$, played the role of space coordinates. 

It was argued by \cite{Euro2} that the large scale properties of the string condensate, within the framework of Loop Quantum Mechanics, are responsible for the effective Riemannian geometry of spacetime at large distances. On the other hand, near Planck scales, the condensate ``evaporates'' and what is left behind is a ``vacuum'' characterized 
by an effective fractal geometry. 

What is required now is to $covariantize$ the Schroedinger Loop wave equation by including the areal time $A$ and the ``spatial coordinates'' $\sigma^{\mu\nu}(C)$ into one single footing. 
We will see that the construction of a $covariant$ Loop Wave functional equation in a $4D$ spacetime, a Klein-Gordon Loop Wave equation,  is attained by introducing a generalized $eleven$ dimensional categorical {\bf C}-space
comprised of $8\times 8$ antisymmetric matrices. The latter matrices encode the generalized coordinates of the histories of points, loops and surfaces $combined$. The points correspond to the center of mass coordinates of a loop/surface. The areal time $A$ and 
$\sigma^{\mu\nu}(C)$ will also  be a part of those     $8\times 8$ antisymmetric 
matrices. We find that  
Spacetime Topology change
and  the Holographic principle are natural consequences of imposing the principle of 
$covariance$ in {\bf C}-space. 
The Planck length must be  introduced  as a necessary rescaling parameter to establish the correspondence limit with the  physics of point-histories  in ordinary Minkowski space,  in the limit $l_P\rightarrow 0$; i.e when the loops and surfaces shrink to a point, the field theory limit.

In section {\bf II} we will review very briefly the Loop Quantum Mechanics of the  Schild-string and write down the Schroedinger Loop Wave equation . In {\bf III} 
we construct boundary wavefunctional solutions to the Loop Wave equation by evaluating the 
phase path integral over open surfaces with boundary $C$.  In {\bf IV}
we propose  an ansatz for  the $SU(\infty)$ Yang-Mills 
vacuum wavefunctional  as a solution of  the Schroedinger Loop Wave equation
associated with  the Loop Quantum Mechanical  formulation of the Eguchi-Schild String . This  String/Loop ( $SU(\infty)$ gauge field) correspondence   
implements one form of the Bulk/Boundary duality conjecture in this case \cite{Mal}.  

Finally in {\bf V} we write down the Covariant Loop Wave Functional equation in the Klein Gordon form ( a classical master field theory) and we argue how Moyal-Fedosov deformation quantization, if , and only if, it is applicable, may be suited to quantize the classical master field theory . 
We will find that the search for a covariant Loop Wave equation where areal time 
$A$ and holographic shadows, $\sigma^{\mu\nu}$,,  are on the same  footing, requires to embed the ordinary string theory in spacetime into the  categorical {\bf C}-space comprised of point, loop and surface histories. All this seems to suggest that the  generalized principle of covariance,  representing invariance of proper $area$ intervals in this {\bf C}-space ( whose elements are antisymmetric matrices) , under matrix-coordinate transformations, 
could be relevant in discovering the underlying principle behind the origins of $M$ theory.

\centerline{\bf 2. LOOP QUANTUM MECHANICS }

We shall present a very cursory description of Eguchi's areal quantization scheme of the Schild string action . For further details we refer to \cite{Euro2}. The starting Lagrangian density  is :

$$L={1\over 4}\{X^\mu, X^\nu\} \{X_\mu, X_\nu\}.~~~X^\mu (\sigma^0, \sigma^1). \eqno (2.1)$$
$X^\mu (\sigma^0, \sigma^1)$ are the embedding string coordinates in spacetime and the brackets represent Poisson brackets w.r.t the $\sigma^0, \sigma^1$ world sheet coordinates. The Schild action is just the $area$ squared of the world sheet instead of the area interval described by the Dirac-Nambu-Goto actions. The corresponding Schild action is only 
invariant under area-preserving transformations . 

Eguchi's areal quantization scheme is based on keeping the area of the string histories $fixed$
in the path integral and then taking the averages over the string tension values.
The Nambu-Goto approach, on the other hand, keeps the tension $fixed$ while averaging over the 
world sheet areas. 
If one wishes to maintain the analog of quantum mechanics of point particles and that of loops, the Eguchi quantization scheme of the Schild action requires the following 
Correspondence principle \cite{Euro2} between area-momentum variables and time ( areal time ) $A$ :

$$P_{\mu\nu} (s) \rightarrow {1\over \sqrt {x'(s)^2 } } {\delta \over \delta \sigma^{\mu\nu}}.
~~~H \rightarrow -i {\partial \over \partial A}. ~~~\hbar =1. \eqno(2.2) $$ 
The 
Loop Schroedinger-like Wave equation $H\Psi =-i(\partial/\partial A)\Psi$ is obtained once we extract the areal time independence , $\Psi =e^{-i{\cal E}A}\Psi$ . 

$$ {1\over l_C} \int_o^1 ds \sqrt {x'(s)^2 } ( { - \delta^2 \over 4 m^2 \delta \sigma^{\mu\nu} \delta \sigma_{\mu\nu} }) \Psi [\sigma^{\mu\nu}(C)]={\cal E} \Psi [\sigma^{\mu\nu}(C)]. 
\eqno (2.3)$$
where the string wave functional   $\Psi [\sigma^{\mu\nu}(C)]$ is the amplitude to find the loop $C$ with area-elements $\sigma^{\mu\nu}$  ( holographic shadows) as the only boundary of a two-surface of internal  area $A$ in a given quantum state $\Psi$.

Plane wave solutions and Gaussian wavepackets ( superposition of fundamental plane wave solutions) were constructed in \cite{Euro2} by freezing all the modes, associated with the arbitary loop shapes,  
except the constant momentum modes on the boundary. The plane wave solutions were of the type :

$$ \Psi [\sigma^{\mu\nu}(C)]\sim exp~[i {\cal E} A-i \oint_C x^\mu (s) dx^\nu (s) P_{\mu\nu}].
\eqno ( 2.4)$$ 
A loop space momentum/string shape-uncertainty principle was given as :

$$\Delta \sigma^{\mu\nu}~\Delta P_{\mu\nu} \ge 1. \eqno (2.5)$$
in suitable units . 

Having presented this very brief review of the Schroedinger Loop Wave equation we shall elaborate further details in the next sections and construct more general solutions than the plane wave case by evaluating the Schild string phase space path integral in {\bf III}. In {\bf IV} we propose  our ansatz for the $SU(\infty)$ YM vacuum wavefunctional as another  candidate solution to the Schroedinger Loop Wave equation. 
And in the final section we present the $covariant$ Loop Wave functional equation and its 
relation to the plausible origins of $M$ theory.   

\centerline{\bf 3. BOUNDARY WAVE FUNCTIONAL EQUATIONS }                                           
We shall construct a boundary wavefunctional solution to the Loop Schroedinger-like wave equation ( besides the standard plane wave solution) which is associated with a Riemann
surface of arbitrary topology and with one boundary. For simplicity we shall start with a Riemann surface of spherical topology,  with the boundary , $\cal C$,  parametrized by the curve  : $x^\mu (s)$. Writing the Schild action in Hamiltonian form allows one to define a boundary wavefunctional solution to the Loop Schroedinger-like wave equation, by  :

$$ \Psi[\sigma^{\mu\nu}(C)]\equiv 
\int_C [D X^{\mu} (\sigma)]
\int  [D P_{\mu a}(s)]
\int_{ \Pi_{\mu a}|_{\partial \Sigma =P_{\mu a }} }~[D \Pi_{\mu a } (\sigma)]
$$
$$ exp~[ i\int {1\over 2}  \Pi_{\mu\nu}d\Sigma^{\mu\nu}(\sigma)
 -i\int d^2\sigma { \Pi_{\mu a}^2\over 4m^2} ]  \eqno (3.1 ) $$
The physical meaning of   $\Psi[\sigma^{\mu\nu}(C)]$ is the probability amplitude of finding any given loop ${\cal C}$ with area-components ( holographic shadows onto the spacetime coordinate planes) $\sigma^{\mu\nu}$ in a given quantum state 
$\Psi$.

One is performing the canonical path integral, firstly, by  summing over all
bulk momentum configurations of the string world sheet  with the restriction that on 
the boundary they equal the pre-assigned value of the boundary momentum, 
$P^{\mu\nu}$ (s), and afterwards ,  one  performs the 
summation  over all values of the $boundary$ momentum, $P^{\mu\nu}$ (s) . 
It is important to emphasize, that one is also summing over all string world sheet embeddings , maps from a world sheet with a disk topolgy to a target spacetime, $X^\mu (\sigma)$ such that maps $X^\mu$ restricted to boundary of the 
disk  are equal to $x^\mu (s)$ . The latter are just the spacetime  parametrization coordinates of 
any  free loop ${\cal C}$; i.e  the boundary shapes , $\cal C$,  are completely $arbitrary$ .

Furthermore, since we are using the Schild action which is not fully reparametrization invariant ( only under area-preserving diffs), we are not performing the functional integral over a family of world sheet auxiliary metrics. 
In \cite{Euro1} we carried  out such functional integral.

The world sheet conjugate area-momentum is defined as the pullback of the target spacetime area conjugate momentum :

$$\Pi^a_{\mu } \equiv\Pi_{\mu\nu} \epsilon^{ab}\partial_b X^\nu (\sigma^a). \eqno (3.2 )$$
and :

$$ \Pi_{\mu a}\Pi^{\mu a}=\Pi_{\mu\nu }\partial_a X^\nu \Pi^{\mu\rho }\partial^a X_\rho = \Pi_{\mu\nu } \Pi^{\mu\rho }\delta^\nu _\rho =          
\Pi_{\mu\nu }\Pi^{\mu\nu }. \eqno (3.3  )$$
which was used in the definition of the Schild Hamiltonian. Where the bulk momentum is restricted to obey the boundary condition : 

$$ \Pi_{\mu\nu} (\sigma^a)|_{\partial \Sigma} = P_{\mu\nu}(s). 
~~~  \Pi_{\mu a}|_{\partial \Sigma} = P_{\mu a}(s)    \eqno (3.4 )$$

The area components enclosed by the arbitray boundary curve $x^\mu (s)$ is given by Stokes theorem :

$$\sigma^{\mu\nu}(C)= {1\over 2} \oint_C (x^\mu dx^\nu- x^\nu dx^\mu)\eqno (3.5  )$$

An integration by parts of the integral  

$$  i\int d^2\sigma \Pi_{\mu\nu}d\Sigma^{\mu\nu}(\sigma) 
 \eqno (3.6a )$$
yields

$$  i\oint_C P_{\mu\nu}d\sigma^{\mu\nu}(s)   
- i\int d^2\sigma X^\mu \{\Pi_{\mu\nu}, X^\nu\}
\eqno (3.6b )                $$
where the Poisson bracket is taken with respect to the world sheet coordinates :$\sigma^0, \sigma^1$. 

The Poisson bracket in the second integrand of eq-(3-6b ) can be written in terms of the canonical conjugate variables : $X^\mu, \Pi _{\mu a}$ because  :

$$  X^\mu \{\Pi_{\mu\nu}, X^\nu\}= X^\mu \epsilon^{ab} \partial_a \Pi_{\mu\nu}(\sigma) 
\partial_b X^\nu (\sigma) =
X^\mu \epsilon^{ab} \partial_a [\Pi_{\mu\nu}(\sigma) 
\partial_b X^\nu (\sigma)]=
X^\mu \partial_a \Pi^a_{\mu  }    \eqno (3.7 )              $$
due to the condition :
$$\epsilon^{ab}\partial_a \partial_b X^\nu =0. \eqno (3.8 )$$
and after using the definition of the world sheet canonical momentum; i.e the pullback of $\Pi_{\mu\nu}$ to the world sheet. Notice that the integrand given by eq-(3-7  ) which is to appear in the exponential weight of the path integral depends solely on $X^\mu $ and $\Pi_{\mu a}$.  
Upon doing firstly the functional integration with respect to all the string embeddings $X^\mu$, of a world sheet of a disk topology into spacetime, one will produce a delta functional constraint which is 
nothing but the classical Schild string equations of motion 
:

$$ \delta (K_\mu ) .~~~ K_\mu=    \{ \{ X_\mu, X_\nu \} , X^\nu \}=\{ \Pi_{\mu\nu} , X^\nu \}=
\partial_a {\Pi}^a_{\mu}=0.\eqno (3.9 )              $$
where we have made use of eqs-(3-1, 3-6b, 3-7, 3.8 ) . 
The on-shell equations of motion are then equivalent to a divergence-free condition on the area-momentum variables  : 

 : 

$$ \partial_a {\Pi}^a_{\mu}=0 . \eqno (3.10 )$$

Therefore, the  $X^\mu$ path integral imposes the $bulk~worldsheet$  classical Schild string equations of motion. This is tantamount of minimizing the proper world sheet area or $areal$  time. A congruence ( or family  ) of classical string configurations is  usualy parametrized by the boundary data  :
$x^\mu (s); P_{\mu a} (s)$. 
This is the string version of the classical free particle motion. One requires to specify the initial position and velocity of the particle to determine the trajectory.

We shall study now  the $quantum$ boundary string dynamics $induced$ by the $classical$ world sheet dynamics. Later, we will extend this procedure to a membrane and all $p$-branes. 
The boundary wavefunctional 
becomes :

$$ \Psi[\sigma^{\mu\nu}(C)]= 
\int  [D P_{\mu a}(s)]~exp~[ i\oint_C {1\over 2}  P_{\mu\nu}d\sigma^{\mu\nu}(s)]
\int_{ \Pi_{\mu a}|_{\partial \Sigma =P_{\mu a }}}~[D \Pi^{class}_{\mu a } (\sigma)]
$$
 $$ exp~ [  -i\int d^2\sigma { (\Pi^{class}_{\mu a}) ^2\over 4m^2} ]  \eqno (3.11 ) $$

The other alternative way to proceed, is firstly by $covariantizing$ the Schild action by introducing auxiliary two-dim metrics, $g^{ab}$ . We explicitly performed such $phase$ space path integral in 
\cite{Euro1} . 
It was also  required  to find the moduli space 
of solutions of the classical Schild string equations of motion,   
$ \partial_a  \Pi^a _{\mu } =0$,  and then to integrate over the moduli.  
Since in two dimensions, a vector is dual to a scalar field, the solutions 
for $\Pi_{\mu a}$ could  be expressed in terms of a set of scalar fields and, in this fashion, we made contact with a bulk Conformal Field Theory. A further functional integral with respect to the auxiliary two-metrics, $g_{ab}$, was also necessary. 
We were able to compute explicitly the bulk momentum path integral and to establish
a $duality$ relation between loop states ( living on ${\cal C}$ ) and string states ( living on the bulk) through a functional loop transform

From this  last expression (3-11) we can recognize immediately that the 
boundary area-wavefunctional $ \Psi[\sigma^{\mu\nu}(C)]$
is nothing but the $loop~transform$ of the boundary momentum wavefunctional which is defined as :

$${\tilde \Psi} [P_{\mu a }] \equiv 
\int_{ P_{\mu a }}[D \Pi _{\mu a } (\sigma)]_{class}~
exp~[ -i\int_\Sigma  d^2\sigma~ [ { (\Pi^{class} _{\mu a})^2\over 4m^2}]~ ]. \eqno (3-12)$$

This bulk momentum path integral defined by eq-(3.12), was $explicitly$ evaluated in 
\cite{Euro1}, when we computed the $covariantized$ Schild action path integral in phase space. We found that the phase space path integral  $factorizes$ into a product of a Eguchi wavefunctional, encoding the 
boundary dynamics, and a bulk path integral term of the Polyakov type, with induced scalar curvature and cosmological constant on the bulk and an induced  boundary cosmological constant and extrinsic curvature on the boundary . The final result of the functional loop transform that maps 
${\tilde \Psi} [P_{\mu a }] $ into   ${\Psi} [\sigma^{\mu \nu }(C)] $ is \cite {Euro1}
:

$$  {\Psi} [\sigma^{\mu \nu }(C)]= \int_0^\infty dA~e^{i\Lambda A} \Psi^{Eguchi} [ \sigma^{\mu \nu }(C); A] 
Z^{Bulk} _A [ \sigma^{\mu \nu }(C)] \eqno (3-13a)$$
and the loop functional was expressed in terms of the holographic coordinates
:$\sigma^{\mu \nu }(C)$ and the loop coordinates $x^\mu (s)$ as follows : 

$$\Psi [x^\mu (s), \sigma^{\mu \nu }(C)]\sim e^{iS_{eff}[C]}{\Psi} [\sigma^{\mu \nu }(C)]. \eqno (3.13b)$$
Notice that $\Psi [x^\mu (s), \sigma^{\mu \nu }(C)]$ is a functional of $two$ arguments and this $differs$ from the standard construction of the 
$\Psi[C]$ in the literature \cite{Gam}. 
$S_{eff}[C]$ is the boundary effective action induced by the quantum fluctuations of the string world sheet. It is a local quantity written in terms of the counterterms needed to cancel  the boundary ultraviolet divergent terms. $\Lambda$ is the world sheet cosmological constant.  $A$ is the world sheet area.    
Such  path integral required to evaluate   the $covariantized$ Schild action, 
$$\int d^2\sigma { \sqrt h}   { (\Pi^{class}_{\mu a}[\sigma; P_{\mu a}(s)])^2\over 4m^2}. \eqno (3.14 ) $$
over each classical momentum trajectory parametrized by the pre-assigned boundary 
momentum,   $P_{\mu a}(s)$, and summing over all possible values of the bulk area of the 
world sheet and auxiliary $h_{mn}$ metrics. In this fashion one eliminates any dependence on the bulk world sheet area. At the end , in order to get    $\Psi [\sigma^{\mu\nu}(\cal C)]$ , one must sum over all the congruence of classical trajectories by integrating over all values of the boundary momentum and auxiliary metrics ( einbeins) induced on the boundary. To regularize ultaviolet infinities requires the introduction of the curvature scalar and cosmological constant on the bulk and the extrinsic curvature and boundary cosmological constant.  

Hence, the boundary area-wavefunctional $\Psi [\sigma^{\mu\nu}(\cal C)]$  associated  with the area-components enclosed by a closed loop  ${\cal C} $ 
is nothing but 
the analog of the Wilson $loop~ transform$  of the boundary momentum wave functional :

$$\Psi [\sigma^{\mu\nu}(C)] \equiv
\int  [D P_{\mu a}(s)]~exp~[ i\oint_C {1\over 2} P_{\mu\nu}(s)d\sigma^{\mu\nu}(s)]~{\tilde \Psi}[P_{\mu a} (s)]. \eqno (3.15  )$$
and viceversa, the inverse Fourier/Loop transform  :

$${\tilde \Psi} [ P_{\mu a} (s)    ] \equiv
\int  [D \sigma^{\mu\nu}] ~exp~[ -i\oint_C {1\over 2} P_{\mu\nu}(s)d\sigma^{\mu\nu}(s)
]~ \Psi [ \sigma^{\mu\nu}]. \eqno (3.16  )$$
yields the boundary momentum wavefunctional  
${\tilde \Psi} [ P_{\mu a} (s)]$ in terms of the 
boundary area-wavefunctional $\Psi [\sigma^{\mu\nu}( C)]$     
with :

$$P_{\mu a} (s) \equiv \Pi_{\mu a}|_{\partial \Sigma}=   (\Pi_{\mu\nu}\partial_a X^\nu)|_{\partial \Sigma}. \eqno (3.17 )$$

It is important   to have  the precise computation of the 
$ {\tilde \Psi} [ P_{\mu a} (s)]$ 
given by eq-(3.12) \cite{Euro1} in order to find  $\Psi [ \sigma^{\mu\nu}(C)]$ as its Fourier/functional loop transform. 
In general, there is an ambiguity in assigning in a unique fashion a string shape ${\cal C}$ to a given set of 
area-components ( Plucker coordinates), $\sigma^{\mu\nu}({\cal C})$. There could be two or more loop-shapes that have the same shadows or coordinates ,  $\sigma^{\mu\nu}$. The Plucker conditions in $D=4$ read :
$$  \epsilon^{\alpha\beta\mu\nu}  \sigma_{ \alpha\beta     } \sigma_{\mu\nu} =0 .\eqno (3.18 )$$ 
to reassure us that there is a one-to-one correspondence between the loop $C$ and its Plucker (area components) coordinates. The classical Schild  action is invariant under $temporal$ area-preserving diffeomorphisms ( it is not fully reparametrization invariant), and one must not confuse this invariance with that of $spatial$ area-preserving diffs : invariance of $(\sigma^{\mu\nu})^2$. One would expect the invariance of the boundary wave functional, under the action of  ( temporal) area-preserving diffs, to  be preserved  in the quantum theory. 
The issue of area-preserving-diffs anomalies will be the subject of future study.

It is straightforward to verify that $\Psi [\sigma^{\mu\nu}(C)]$ defined by eq-( 3-15  ), is a solution of the 
Loop Schroedinger-like Wave equation $H\Psi =-i(\partial/\partial A)\Psi$ once we extract the areal time independence , $\Psi =e^{-i{\cal E}A}\Psi$ . 

$$ {1\over l_C} \int_o^1 ds \sqrt {x'(s)^2 } ( { - \delta^2 \over 4 m^2 \delta \sigma^{\mu\nu} \delta \sigma_{\mu\nu} }) \Psi [\sigma^{\mu\nu}(C)]={\cal E} \Psi [\sigma^{\mu\nu}(C)]. \eqno (3.19 )$$
where the on-shell dispersion relation for the Eguchi-Schild string is :

$${\cal E} ={1 \over l_C} \int_0^1  ds \sqrt {x'(s)^2 }  
{P_{\mu a}^{class}(s) P^{\mu a}_{class} (s) \over 4m^2 }. \eqno (3.20 )$$
with $l_C$ being the reparametrization invariant length of the loop, ${\cal C}$. 

The expectation value of the Hamiltonian Operator,  in the state 
$\Psi [\sigma^{\mu\nu} ({\cal C})]$, after inserting the expression  ( 3-15   ),  and using the definition of $\delta [ P_{\mu a}-
P'_{\mu a}]$, 
is simply :

$$ <~{\cal H}~>_{\Psi}= {1\over l_C}  \int_0^1 ds \sqrt {x'(s)^2 }\int [D\sigma^{\mu\nu} (s)] \Psi^* [\sigma^{\mu\nu}(C)]
( { - \delta^2 \over 4 m^2 \delta \sigma^{\mu\nu} \delta \sigma_{\mu\nu} } )  
\Psi [\sigma^{\mu\nu}(C)] =$$

$${1\over l_C} \int_0^1 ds \sqrt {x'(s)^2 }~\{ \int [D P_{\mu a} (s)] 
{\tilde \Psi}^*[P_{\mu a} (s)] 
{P_{\mu a}(s) P^{\mu a} (s) \over 4m^2 } {\tilde \Psi}[P_{\mu a} (s)]~\}= $$

$${1\over l_C} \int_0^1 ds \sqrt {x'(s)^2 } <~{P_{\mu a}(s) P^{\mu a} (s) \over 4m^2 }~>_{\Psi (P)}  =
{\cal E}. \eqno (3.21  )$$
where the quantity under the brackets in eq-(3.21) is the defining expression for the 
averages of $ P_{\mu a}(s) P^{\mu a} (s) \over 4m^2$ over the boundary momentum quantum states. 
As expected we recover the on-shell dispersion relation for the Eguchi-Schild string if, and only if, 
$$<~{P_{\mu a}(s) P^{\mu a} (s) \over 4m^2 }~>_{\Psi (P)}=
{P_{\mu a}^{class}(s) P^{\mu a}_{class} (s) \over 4m^2 }. \eqno (3.22  )$$
This is precisely what is expected for a $free$ string; i.e eigenstates of the boundary momentum operator, ${\hat P}_{\mu a} (s)$. It is important also to impose the normalization condition :

$$\int [D P_{\mu a} (s)] {\tilde \Psi}^*[P_{\mu a} (s)] {\tilde \Psi}[P_{\mu a} (s)]=1. 
\eqno (3.23) $$ 
Concluding , $\Psi [\sigma^{\mu\nu}({\cal C} )]$ solves the Loop Schroedinger-like wave equation and the expectation value of the Hamiltonian operator is indeed equal to those values of ${\cal E}$ 
which are $consistent$ with the Eguchi-Schild string on-shell dispersion relation. 

For a membrane of spherical  topology spanning a world tube $S^2\times R$ the higher-loop analog of the loop wave equation is : 
$$ {1\over A} \int d^2\sigma {\sqrt g}  {1\over 2.3! m^3}  
( { - \delta^2 \over  \delta \sigma^{\mu\nu\rho}(S^2) \delta \sigma_{\mu\nu\rho}(S^2)  }) \Psi [\sigma^{\mu\nu\rho}(S^2)]={\cal E} \Psi [\sigma^{\mu\nu\rho}(S^2)]. \eqno (3.24 )$$
where $ \sigma^{\mu\nu\rho}(S^2)$ will be the volume components enclosed by the spherical membrane ( bubble) $S^2$; i.e the proyection (shadows) of the volume onto the target spacetime coordinate planes. This  higher-loop boundary wave equation corresponds to the quantum dynamics of a Euclidean world sheet or bubble, the sphere $S^2$; i.e the boundary of the world tube of the membrane . 
The ${\cal E}$ eigenvalue has units of membrane tension : energy per unit area 
($mass^3$). The $A$ represents the reparametrization invariant area of the bubble (sphere).  The action for the membrane is in this case the generalization of the Schild action, a volume squared 

$${1\over l^3}\int d^3\sigma 
{\partial (X^\mu, X^\nu, X^\rho)\over \partial (\sigma^0,\sigma^1, \sigma^2)}^2. \eqno (3.25 )$$ 

The Mechanics associated with p-branes admits the so-called Nambu-Poisson Hamiltonian Mechanics where the ordinary Poisson bracket of two quantities 
$\{A, B\}$ is replaced by a multi-bracket : $\{A,B,C....\}$ which is essentially the volume form/Jacobian. 
The subject of Higher-Dimensional Loop Spaces and their algebras is a very complex one. We refer to \cite{Ced} for an introduction.

\centerline {\bf 4. Wavefunctionals of Loops and $SU(\infty)$ Gauge Theories }

In this section we shall construct loop wavefunctionals, $\Psi [{\cal C}]$ associated with a given loop and must not be confused with area-wavefunctionals of the previous section :
$\Psi [\sigma^{\mu\nu} ({\cal C})]$ associated with a given loop
${\cal C}$ with area-components $\sigma^{\mu\nu} ({\cal C})$.   

Based on the known observation \cite{Fai}, \cite{Hop} that classical ``vacuum'' configurations of a
$SU(\infty)$ YM theory ( space time $independent$ gauge field configurations) 
are  given by classical solutions of the Eguchi-Schild string, after the 
$A^\mu \leftrightarrow X^\mu (\sigma^0, \sigma^1)  $ correspondence is made, we shall construct ``vacuum'' loop functionals $\Psi_{vac} [\gamma]$ 
using the standard  Wilson loop transform 
associated with the $SU(\infty)$ YM theory; i.e gauge theory of area-preserving diffs. 
By ``vacuum'' one means classical solutions to the    $SU(\infty)$ YM equations of motion 
for the special case when the $A^\mu (x^\mu;\sigma^a)$  fields are spacetime $independent$ :

$$D_\mu F^{\mu\nu}= \partial_\mu F^{\mu\nu} + \{A_\mu , F^{\mu\nu} \}=0 
\Rightarrow  \{A_\mu , \{A^\mu, A^\nu\}\}=0 \leftrightarrow 
\{X_\mu , \{X^\mu, X^\nu\}\}=0. \eqno (4.1 )$$
after the gauge field/string correspondence is made one recovers the classical Schild string equations of motion. From now on we shall use the term ``vacuum'' to represent the classical solutions of the Schild string and must not be confused with the true vacuum of the theory; i.e the state which has true zero 
expectation values for all physical observables :$ <F^{\mu \nu}>_{vac}=0$ and  zero energy,  $ <H>=<F_{\mu \nu} F^{\mu \nu}>_{vac}=0$.   

The Wilson loop holonomy operator associated with a $SU(\infty)$ gauge field , $A^\mu (x^\mu, \sigma^0, \sigma^1)$  is defined :  

$$W[A^\mu, \gamma ]=Tr~P~exp ~[i\oint_\gamma A_\mu dx^\mu ]_{N\rightarrow \infty} = 
\int d^2\sigma exp ~[i\oint_\gamma A_\mu dx^\mu ]=
<\gamma|A^\mu>     \eqno (4.2  )$$
and its complex conjugate associated with a different contour ${\cal C}$ is :

$$W^*[A^\mu, C ]=  Tr~P~exp~ [-i\oint_C A_\mu dx^\mu ]_{N\rightarrow \infty} = 
\int d^2\sigma exp ~ [-i\oint_C A_\mu dx^\mu ]= <A^\mu |x^\mu (s)>  .\eqno (4.3 )$$
we choose the contours $\gamma, {\cal C}$ to be different for reasons which will become clear later. .  
 For the time being we shall not be concerned with path orderings nor traces. After all,  the $SU(\infty)$ gauge field is now an ordinary number and not a matrix. The trace in the $SU(\infty)$ YM case is replaced by an integral w.r.t the internal space ``color''  $\sigma^a$ coordinates. 

There is a signature subtlety if one wishes to identify the $\sigma$ internal color coordinates with the string worldsheet ones. The former live in a Euclidean internal world whereas the latter in a Minkowskian spacetime. However, if one performs a dimensional reduction of the $A^\mu$ field to one temporal dimension ( Matrix Models \cite{Ban}) one makes contact with a true physical membrane coordinate :
$A^\mu (t, \sigma^0, \sigma^1)\leftrightarrow X^\mu ( t, \sigma^0, \sigma^1)$ 
because now we have the correct signature for the world volume of the membrane. A timelike slice of the membrane reproduces a string worldsheet whereas a spatial slice a Euclidean  worldsheet. The signature subtleties have been recently studied by \cite{ Hul}  and \cite{Amb} .

We turn attention to what we consider one of the important aspects in the essence of duality : $q\leftrightarrow p$ in the standard Hamiltonian dynamics. It has been a long-sought goal to construct a formulation of extended objects where dualities are already manifest. A formulation of all bosonic p-branes as Composite Antisymmetric Tensor Gauge theories of the volume-preserving diffs where $S,T$ duality were incorporated from the start was achieved  in \cite{Cas3}. 
For example, in the standard canonical quantization of electromagnetism , ${\vec A}, {\vec E}=F_{oi}$ are a canonical pair of conjugate variables. In the Schild formulation one has : $X^\mu, \Pi_{\mu a}$ as a canonical pair which have  a correspondence with  the $SU(\infty)$ $c$-number variables : 
$A^\mu, F_{\mu\nu}$, respectively .The $c$-number valued field strength is :

$$F_{\mu\nu}=\partial_\mu A_\nu -\partial_\nu  A_\mu + \{ A_\mu, A_\nu \}_{PB}. \eqno (4.4 )$$
the Poisson brackets are taken w.r.t the internal color indices : $\sigma^0, \sigma^1$. 
Moyal deformations are very natural, for recent results on this we refer to 
\cite{Kav}, \cite{Gra}
, \cite{Cas2}. 
For ( vacuum solutions ) spacetime $independent$ field configurations one can make direct contact with the Schild string 
string  since the classical YM Lagrangian becomes now equivalent to the Schild one :

 $$F_{\mu\nu}^2=\{ A_\mu, A_\nu \}_{PB}\{ A^\mu, A^\nu \}_{PB}\leftrightarrow
\{ X_\mu, X_\nu \}_{PB}\{ X^\mu, X^\nu \}_{PB}    \eqno (4.5 )$$

Spacetime independent field configurations yield  a trivial holonomy factor ( since the vector sum of all the tangents along a closed loop is zero)  leaving 
only the internal color space $area$ operator :
$ W[A,\gamma]=Area [\int d^2 \sigma]$ as the holonomy. If one  wishes to identify the colour coordinates with the world-sheet ones, colour confinement would imply a finite colour-area : a compact world sheet with boundary  
$C$ that will be mapped into the Wilson loop $\gamma$ living in the target spacetime.   
It is important to remark that if one evaluates the loop derivative on the Wilson loop
$W[A^\mu,\gamma]$ in the vacuum case, one must $firstly$ take the derivatives and $afterwards$  set the values of $A^\mu =A^\mu (\sigma^a)$ yielding :

$${\delta \over \delta x^\mu (s)}W[A^\mu,\gamma]= x'^\nu F_{\mu\nu}(x^\mu (s)) W[A^\mu, \gamma].~~~x'^\nu (s)={\partial x^\nu (s)\over \partial s}. \eqno (4.6 )$$
otherwise one will trivially get zero since the spacetime loop derivative does not affect the area in the internal colour space.

These  area and volume operators are very important in Loop Quantum Gravity \cite{Gam}  especially in regards to the fact that these operators have $discrete$ eigenvalues : spacetime is quantized in multiples of Planck areas and Planck volumes. This will 
not be surprising if the group of area and volume preserving diffs are truly 
relevant symmetry groups in nature. However, it is important to realize that area in color space is not the same as area in spacetime. Strings and membranes can be interpreted as gauge theories of the area-preserving diffs algebra,\cite{Sug} where the effective dimensions will be 
$D+2$, two internal dimensions ( the infinite color space) in addition to the usual ones.   
If one identifies the internal color directions, $\sigma^0, \sigma^1$,  with an Euclidean worldsheet one makes contact with the Euclideanized string for those spacetime independent gauge field configurations of the $SU(\infty)$ gauge field.  In the case that 
$A^\mu (x^\mu;\sigma^a)$ is dimensionally reduced to one temporal dimension one will then 
make contact with the membrane ( this time with the correct signature); i.e with M(atrix) Models.

Therefore, the  $SU(\infty)$ gauge field/Schild string correspondence  is then :

$$W[A^\mu,\gamma] = \int d^2\sigma exp ~[i\oint_\gamma  A_\mu dx^\mu 
 ] \leftrightarrow exp ~[i\int \Pi_{\mu\nu}  d\sigma^{\mu\nu}]=$$
$$exp ~[i\oint_C P_{\mu\nu}(s)  d\sigma^{\mu\nu}(s)  -i          
\int d^2\sigma X^\mu \{ \Pi_{\mu\nu}, X^\nu \}]. \eqno (4.7 )$$
after an integration by parts. The Eguchi-Schild string equations of motion set to zero the last term of (4-7) and one gets ( at the classical vacuum level) the correspondence  :                                                                        
$$W[A^\mu,\gamma] = Area  [\int d^2 \sigma]    \leftrightarrow 
exp ~[i\oint P_{\mu\nu}(s)  d\sigma^{\mu\nu}(s)]. \eqno (4.8  )$$
The correspondence between the loop transform of $\Psi [A^\mu]$ and that of the
loop momentum 
boundary wavefunctional of the Eguchi-Schild string is :
$$\Psi [\gamma] \leftrightarrow 
  \Psi [\sigma^{\mu\nu}(C) ]$$ 
$$\int [DA^\mu]W[A^\mu,\gamma]\Psi [A^\mu]  \leftrightarrow 
\int [DP_{\mu a}(s)] exp ~[i\oint_C P_{\mu\nu}(s)  d\sigma^{\mu\nu}(s)]\Psi [P_{\mu a} (s)]. \eqno (4.9 )$$ 

We emphasize once more that  $\Psi [\gamma]$  and   
$\Psi [\sigma^{\mu\nu}(C) ]$ are $not$ the same  objects even in the case that $\gamma={\cal C}$. There is a $correspondence$ between them. 
 $\Psi [\gamma]$ should obey a suitable $SU(\infty)$ loop equation and  
$\Psi [\sigma^{\mu\nu}(C) ]$ obeys the Loop Schroedinger-like Wave equation discussed in the previous section. We propose that $\Psi [\gamma]$ ( for the vacuum) should obey the following Loop
Wave equation \cite{Euro2}  :

$${1\over l_C} \int_0^1 {ds\over \sqrt {x'(s)^2}} ( {1\over 4m^2}  {-\delta^2 \over 
\delta x^\mu (s) \delta x_\mu (s)}) \Psi_{vac} [\gamma]={\cal E} \Psi_{vac} [\gamma] . \eqno (4.10 )$$
Notice that this Loop wave equation clearly differs from the $SU(N)$ loop wave equation in the literature \cite{Gam} . In the $N\rightarrow \infty$ limit, the standard  loop equation would require a wavefunctional of an $infinite$  number of loops
$\Psi [\gamma_1,\gamma_2.........]$ . Whereas, the Loop equation above (4.10  ) requires one loop only. It can be generalized to many loops (closed strings) when interactions are included. What allows us to avoid the problem of using an infinite number of loops is precisely when we impose the  spacetime independent ( vacuum) $SU(\infty)$ YM gauge field configurations /
Eguchi-Schild string correspondence.        

Such gauge field/string  \cite{Pol1}, \cite{Pol2} correspondence has  been discussed by many authors in particular by 
\cite{Bae1}  pertaing strings, loops, knots, quantum gravity and BF theories and also by 
\cite{Gro} . The former authors have shown that the large $N$ limit of QCD in $D=2$  corresponds to 
$a ~string $ theory, although it was never established $which$ particular string theory it actually referred to. The action for the string theory that reproduces the partition function of the large $N$ QCD in $D=2$ was never constructed. The quantum YM partition function could be matched with an infinite sum of branched string coverings ( with singularities) of a family of Riemann surfaces  of arbitary genus, $g$ to a compact Riemann surface of fixed genus $G$ and fixed area. The latter is the compact Riemann surface where the orginal $SU(\infty)$ YM field theory lived in.

Hence, we propose that the Wilson loop transform of the vacuum $SU(\infty)$ YM wavefunctional, $\Psi_{vac}[A^\mu] \rightarrow   \Psi_{vac}[\gamma]$ is a $solution$ of the Schild string Loop Wave equation described by eq-( 4-10 ). Furthermore, due to the $A^\mu \leftrightarrow X^\mu$ correspondence, the $\Psi_{vac}[\gamma]$ can also be obtained by performing the Schild string path integral over a Riemann surface of spherical topology with one boundary equal to the closed loop   $\gamma$ :

$$ \int_{SU(\infty)}  [DA^\mu]W[A^\mu, \gamma] \Psi_{vac}[A^\mu]= \Psi_{vac}[\gamma]= 
\int_{X^\mu |\partial \Sigma =\gamma}[DX^\mu (\sigma)] e^{iS_{Schild}[X^\mu]}. \eqno (4.11 )$$
the family of maps , $X^\mu (\sigma)$ from a Riemann surface of a disk topolgy to 
spacetime  is restricted to obey a boundary condition : the surface must have as boundary 
the  loop $\gamma$.

The string path integral can be evaluated  $perturbatively$ ( approximately). However, one will not recapture the nonperturbative string physics in that fashion. String perturbative corrections are obtained as usual by summing over all surfaces of arbitrary genus. The path integral can be evaluated in different steps :
we fix first the genus and area, then sum over all areas and afterwards over all genus. Auxiliary metrics are introduced in the $covariantized$ Schild action. The latter is on-shell equivalent to the Nambu-Dirac-Goto and Polyakov actions. A sum over all metrics is then performed and in this way the path integral is computed modulo the action of the diffs and Weyl group. At the end one ends up with integrals in the Moduli space ${\cal M}_{h,1}$ of 
Riemann surfaces with $h$ handles and one puncture ( associated with one boundary). The invariant measure in the moduli space is obtained by means of the $b,c$ ghost contributions ( Faddev-Popov determinants). Noncritical string dimensions require the Liouville field.

Inserting $\Psi_{vac}[\gamma]$ provided by the standard Wilson loop transform 
, left hand side of (4-11 ),  into the Loop Wave equation (4-10  )  yields for the expectation values of the YM Hamiltonian ( that corresponds to the Schild Hamiltonian) after using eq-(4-5  ) :

$${1\over l_C} \int_0^1 ds \sqrt {x'(s)^2}  {1\over 4m^2}\int [DA^\mu]\Psi_{vac}^*[A^\mu]
~(g^2_{YM} {F_{\mu\nu}^2[x^\mu (s)]\over  g^2_{YM}  } )~ \Psi_{vac}[A^\mu] ={\cal E}_{vac}.\eqno (4.12  )$$
and one  obtains that the $vacuum$ expectation value of the $SU(\infty)$ YM energy 
density, $ (F_{\mu\nu}^2[x^\mu (s)]/g^2_{YM}  )$, restricted to the values on the loop $x^\mu (s)$,  is precisely proportional to the Schild string tension.  This is due to the fact  that ${\cal E}$ has units of energy per 
unit length which is a tension .    
We have introduced  the YM coupling , $g_{YM}$ in the form 
$g^2_{YM} (F_{\mu\nu}^2/ g^2_{YM})     $ to match units appropriately. 
The latter term has units of 
$length^{-4}$ as it should in order that the l.h.s of (4.12) has the units of a string tension. For recent work on the QCD vacuum wavefunctional and the large $N$ 't Hooft's expansion in relation to strings see \cite{Bro}, \cite{Kak}.

 Inserting $\Psi_{vac}[\gamma]$ provided by the right hand side of ( 4-11 ) yields a solution of the Loop Wave equation after one identifies the 
$\Psi_{vac}[\gamma]$ ( up to a  normalization constant $N$ of units $L^{-1/2}$ to absorb the units coming from the 
$X^\mu$ integral ) with the  boundary effective action $induced$ by the bulk quantum dynamics :  
$$  \Psi_{vac}[\gamma]=e^{iS_{eff}[x^\mu (s)]} \equiv \int_{X^\mu |\partial \Sigma =\gamma}[DX^\mu (\sigma)] e^{iS_{Schild}[X^\mu]}. \eqno (4.13 )$$
In ordinary $SU(2)$ YM gauge theories in $D=4$ , the Chern-Simmons wavefunctional $\Psi [A] =e^{iS_{CS}[A]}$ ( the boundary term)  is used to evaluate the loop transform from $\Psi [A]\rightarrow \Psi [\gamma]$. In the $SU(\infty)$ case we shall use the boundary effective action induced by bulk Schild string quantum dynamics. 

Plugging the above solution (4-13) , into the Loop Wave equation yields for the expectation value of the Schild Hamiltonian density :

$${1\over l_C} \int_0^1 {ds \over \sqrt {x'(s)^2}}  {1\over 4m^2}\int [D x^\mu(s)]N^2~
 [  {  \delta^2 S_{eff}[x^\mu (s)]\over \delta x^\mu \delta x_\mu} +      
({\delta S_{eff} [x^\mu (s)]\over \delta x^\mu})^2 ]={\cal E}_{vac}.\eqno (4.14  )$$
The $({\delta S_{eff} [x^\mu (s)]\over \delta x^\mu})^2$ are the standard momentum squared terms ( to order $\hbar^0$), 
kinetic terms  of  the Hamilton-Jacobi equation and the  
${  \delta^2 S_{eff}[x^\mu (s)]\over \delta x^\mu \delta x_\mu}$ are 
the so-called  ``quantum potential `` terms ( to order $\hbar$).    
We can see that the units match appropriately.

Given a $\Psi_{vac}[\gamma]$ associated with the $SU(\infty)$ YM  theory that obeys the Loop wave equation, the next question to ask : is what happens for states $other$ than the vacuum ? Clearly one cannot hope to use the Eguchi-Schild correspondence any
longer.   
We used, originally,  the Schild string as a guiding principle but one should $not$ expect the Schild string to be the actual theory behind the full fledged   
$SU(\infty)$ YM theory, especially in $D \ge 2$.    
 At the classical level there was a simple correspondence between the Schild vacua and those vacua of the $SU(\infty)$ YM theory ( space time independent field configurations). However, this is a very $restricted$ case.  The question is again : what $string$ theory, if any, can one use to construct wavefunctionals other than the vacuum in any dimension ? 

Since $W_{\infty}$ is an area-preserving diffs algebra \cite{Bou}  , containing the Virasoro algebra,  which appears naturally in the physics of membranes, 
we believe that the sort of string theory one is looking for could be a $W_{\infty}$ string theory; i.e an Extended Higher Conformal Spin Field Theory representing $W_\infty$  
gravity coupled to $W_{\infty}$ Conformal Matter plus an infinite tower of Liouville fields and ghosts in the noncritical $W_\infty$ string case . It has been  formally shown in \cite{Cas3}  that $D=27/11$ are the 
appropriate target spacetime dimensions for the bosonic/supersymmetric non-critical $W_\infty$ string,  if the theory is devoid of BRST anomalies. It is interesting to see that these  are precisely the membrane/supermembrane spacetime dimensions that are devoid of Lorentz anomalies \cite {Mar}  . The self dual sector of the membrane/supermembrane spectrum contains  non-critical   $W_{\infty}$ strings/superstrings.   Higher Spin Gauge theories have been revisited very recently by 
\cite{Vas}, \cite{Sez} in connection to Higher Spin Gauge interactions of Massive fields in 
Anti-de-Sitter space in $D=3$ and $N=8$ Higher Spin Supergravities. It has been argued that 
Higher Spin Gauge theories may actually be the underlying theory behind the bulk dynamics of 
Anti-de-Sitter space in any dimensions. For a review of the historical precedents of Maldacena's conjecture \cite{Mal}
we refer to Duff \cite{Duf}.

In the case that the boundary loops do $not$ coincide, $\gamma$ is not equal to ${\cal C}$, we propose the following  vacuum loop wavefunctional  as the solution to the Loop Wave equation. In \cite {Euro1}  we constructed   the area-wavefunctional $\Psi [\sigma^{\mu\nu}(\gamma)]$ 
directly from the computation of $covariantized$ Schild string phase space path integral. 
The latter is $not$ the same as the $\Psi[\gamma]$ defined as
 :

$$\Psi_{vac}  [\gamma] =<\gamma|\Psi'_\Sigma>=
\int [DA^\mu]<\gamma|A^\mu> <A^\mu|\Psi'_\Sigma>=$$
$$\int [DA^\mu]  W[A^\mu,\gamma ]\int [Dx^\mu (s)]  <A^\mu|x^\mu>
\int_0^\infty  dAe^{i{\cal E} A (\gamma,C) }<x^\mu |\Psi_\Sigma> =$$
$$ \int [DA^\mu]  W[A^\mu, \gamma ]\int [Dx^\mu (s)]
W^* [A^\mu, C]  \int_0^\infty  dA e^{i{\cal E} A (\gamma,C)}\Psi_\Sigma [x^\mu (s)]. 
\eqno (4.15  )$$
where :
$A=A(\gamma, C)$ is the proper area whose boundaries are $\gamma, C$; i.e. Areal time spanned between $\gamma,C$.  
We have used the relation $|\Psi'_\Sigma>= e^{i{\hat H}A}|\Psi_\Sigma>=e^{i{\cal E} A}|\Psi_\Sigma>$
since the Schild Hamiltonian operator is the evolution operator of the initial quantum eigenstate $|\Psi_\Sigma>$ ( assigned to the boundary $C$) ,  to the final  quantum eigenstate  $|\Psi'_\Sigma>$ assigned to the boundary $\gamma$,   
in a given areal time : $A(\gamma, C)$. At the end one must sum over all possible values of the proper area spanning between $\gamma$ and $C$.  

The wavefunctional assigned  to the $C$ boundary is once again : 
$$\Psi_\Sigma [x^\mu (s)]\equiv \int_{x^\mu (s)}  [DX^\mu (\sigma)]
exp~[i S_{Schild}[X^\mu (\sigma)] ] = e^{i S_{eff}[x^\mu (s)]}.           \eqno (4.16 )$$
The path integral of the bulk field theory is evaluated on all the open surfaces that have 
the loop ${\cal C} $ as the boundary  : $X^\mu (\sigma)|_{\partial \Sigma} =x^\mu (s)$. 
The wavefunctional $\Psi_\Sigma [x^\mu (s)]$ agrees exactly with the one described by the operator formalism  of Riemann surfaces \cite{Alv} . The Schild action functional, 
$S_{Schild} [X^\mu]$,  will reassure us that 
$\Psi_{vac}[\gamma] $ will indeed be a solution of the 
Loop wave equation (4-10  ) if, and only if,  $\Psi_\Sigma [x^\mu (s)]$ obeys the Loop Wave equation as it should, since it was constructed by performing the Schild-string path integral. 
Therefore, the two wavefunctionals must be related through the Schild-string kernel :

$$\Psi_{vac} [\gamma]= \int [Dx^\mu (s)]\int_0^\infty dA~ K_{Schild}[C,\gamma, A(\gamma,C) ]   \Psi_\Sigma [x^\mu (s)]. \eqno (4.17 )$$
with $A[\gamma, C]$ being the proper $area$ ( areal time)  spanned between the initial loop $C$ and the final loop $\gamma$.                                      
In the liming case that the two loops coincide ( $A=0$)  the Schild loop propagator turns into a delta functional $\delta [\gamma - C]$ and one gets from eqs- (4-11, 4-17 ) :

$$\Psi_\Sigma [x^\mu (s)]=\Psi_{vac} [\gamma]= \int [DA^\mu]  W[A^\mu, C ] \Psi_{vac} [A^\mu] \eqno (4.18)$$ 
which is indeed the $loop~transform$ of the vacuum wavefunctional of the $SU(\infty)$ YM theory. And the latter, in turn, is $indirectly$ given in  terms of the $bulk$ field theory by using, precisely,  the $inverse$ loop transform of 
the boundary wavefunctional associated with the conformal field theory living 
on the $bulk$; i.e the Riemann surface ( string world sheet) with boundary 
$\gamma= {\cal C}$   :

$$\Psi_{vac}[A^\mu]= \int [Dx^\mu (s)] W^* [A^\mu, \gamma] \Psi_{vac} [\gamma]. \eqno ( 4.19 )$$

Concluding : It is precisely when one sets $\Psi_{vac}[\gamma]= \Psi_\Sigma [x^\mu (s)=C]$, that we have the following 
string (bulk)/$SU(\infty)$ gauge field (boundary) duality for the vacuum wavefunctionals:

$$\Psi_{vac}[\gamma]=\int [DA^\mu]  W[A^\mu, \gamma ] \Psi_{vac}[A^\mu]= \int_{x^\mu (s)=\gamma}  [DX^\mu (\sigma)]
exp~[i S_{Schild }[X^\mu (\sigma)] ] . \eqno (4.20 )$$
the left hand side is just the Wilson loop transform of the $SU(\infty)$ YM vacuum wavefunctional whereas the right hand side is the induced boundary effective action   
defined above by performing the path integral of the Schild string bulk field theory with the 
standard boundary restriction on the string embeddings : $X^\mu |_{\partial \Sigma} =x^\mu (s)=\gamma$.  

For states $other$  than the vacuum we postulate that :

$$\Psi[\gamma]=\int [DA^\mu]  W[A^\mu, \gamma ] \Psi [A^\mu]= \int_{x^\mu (s)=\gamma}  [DX^\mu (\sigma)]
exp~[i S_{bulk }[X^\mu (\sigma)] ] . \eqno (4.21 )$$

The $crucial ~question$  is once more  : What is the theory behind $S_{Bulk}[X^\mu]$ in the more general case besides the vacuum ? Is it a Higher Spin Gauge Theory; i.e a 
$W_{\infty}$ gauge theory ? A $W_{\infty}$ string theory ? As stated earlier, we used, originally,  the Schild string as a guiding principle but one should $not$ expect the Schild string to be the actual theory behind the action functional $S_{Bulk}[X^\mu]$.  
We believe that $W_{\infty}$ geometry, higher spin gauge theories ,  could be  the theory behind the $S_{Bulk}[X^\mu]$. $W_{\infty}$ algebras and their Moyal \cite{Moy}, \cite{Fed}  
and $q$ deformations are the natural candidates to construct the Quantum Gauge Theories of the area-preserving diffs group.

To conclude, string/gauge field duality is tantamount of a bulk/boundary duality for vacuum 
$SU(\infty)$ gauge field configurations when $\gamma={\cal C}$.  
In the more general case that $\gamma$ dose not coincide with $ {\cal C}$ 
the Schild loop propagator $K[C,\gamma,A]$ is the kernel operator that evolves the initial state  $\Psi_\Sigma [C]$, 
in a given   areal time $A(\gamma, C)$
 between the two loops $\gamma, {\cal C}$,  to a new quantum state  
$\Psi_{vac} [\gamma]$. This is equivalent to attaching a plumbing fixture or throat , typical of Closed String Field Theory \cite {Bar}, to the original Riemann surface ( 
disk topology ) at the boundary $C$ extending it all the way to the final boundary
$\gamma$. And then, propagating ( sliding ) the intial loop ${\cal C}$ along 
the throat all the way to the final loop $\gamma$. As it propagates, the shape of the intitial loop ${\cal C}$ changes ( ``elastic'' ) to match $\gamma$ at the other end. And, finally, one performs the sum over all values of areal time ( area) $A$ as shown in (4-15 ). 

This is is consitent with the fact that if $\Psi_\Sigma [C],  \Psi [\gamma]$ 
are  both solutions of the loop wave equation, then 
the standard operator form of the Loop wave equation gives  : 
$${\hat H}|\Psi>= {\cal E}|\Psi>~~~
    e^{i {\hat H} A(\gamma, C)} | \Psi>= e^{i {\cal E} A(\gamma, C)} | \Psi>= | \Psi'>   . \eqno (4.22  ) $$
It is clear that when $\gamma \rightarrow {\cal C} $ the area ( areal time)  tends to zero 
and both wavefunctionals will coincide as expected. 
Finally, the induced boundary effective action was required  to be identified with the phase of the boundary vacuum   
wavefunctional $  \Psi_{vac} =\Psi_\Sigma [C]=Nexp[iS_{eff}[x^\mu (s)]]$ if this boundary wavefunctional 
is  a true solution of the Vacuum Boundary Loop Wave equation . 

To finalize this section, we emphasize once more that one must differentiate between the three different representations associated with the quantum state $|\Psi>$. The 
$\Psi_{vac} [\sigma^{\mu\nu} (\gamma)]$, is the area-wavefunctional that was evaluated explicitly in \cite{Euro1}, and is related to   
the standard  loop functional $\Psi_{vac} [\gamma]$ as  follows : 

$$\Psi [\sigma^{\mu\nu} (\gamma)]=< \sigma^{\mu\nu}|\Psi>= \int [D\gamma]   
< \sigma^{\mu\nu}|\gamma>  < \gamma|\Psi> = \int [D\gamma]   
< \sigma^{\mu\nu}|\gamma> \Psi [\gamma] . \eqno (4.23  ) $$ 
 
The quantity $< \sigma^{\mu\nu}|\gamma>$ represents the probability amplitude that the loop $\gamma$ has for Plucker area-components, the values of $ \sigma^{\mu\nu}$. As usual, 
$\Psi [\gamma]$ is the loop wavefunctional discussed previously in full detail; i.e the Wilson loop transform assocated with the $SU(\infty)$ YM vacuum functional $\Psi_{vac} [A^\mu]$ or defined via the string path integral with boundary $\gamma$. 

The Fourier conjugate state given earlier by the functional loop transform in eqs-(3.15,3.16) can also be represented as   :
$${\tilde \Psi}_{vac}  [P_{\mu a}(s)    ]=     <  P_{\mu a}|\Psi>= \int [D\gamma]   
<    P_{\mu a}|\gamma>  < \gamma|\Psi> = \int [D\gamma]   
<   P_{\mu a} |\gamma> \Psi_{vac} [\gamma] . \eqno (4.24  ) $$ 
and coincides with the expression (3.12).  $ {\tilde \Psi}  [P_{\mu a}(s)    ]$. 

The quantity $<   P_{\mu a} |\gamma>$ represents now the probability amplitude that the given loop $\gamma$ carries an area-momentum of  $ P_{\mu a}(s)$.

Concluding, finally, the three descriptions : 
$$  \Psi_{vac} [\gamma],~\Psi_{vac} [\sigma^{\mu\nu} (\gamma)], ~ {\tilde \Psi}_{vac}  [P_{\mu a}(s)    ]. \eqno (4-25  )$$ 
are nothing but just three different representations of the $same$ quantum state, $|\Psi>_{vac}$ of 
the $SU(\infty)$ YM theory : the Wilson loop, the area  and the momentum  
representations, respectively. Since the area-wavefunctional representation 
$\Psi_{vac} [\sigma^{\mu\nu} (\gamma)]$ was given explicitly by \cite{Euro1} one can then evaluate, in principle, the functional maps which take one representation into the other.

 \centerline {\bf 5  The Master Covariant Loop Wavefunctional Equation 
and M Theory}

The loop wave equation that has been discussed so far are of the Schroedinger type. This is  not covariant since $A$ ( areal time)  and $\sigma^{\mu\nu}$ ( the analog space like coordinates) are on a  different footing. What we need now is the
analog of the Klein Gordon equation. Pavsic \cite {Pav} has written a covariant functional Schroedinger equation in the Fock-Schwinger proper time formalism by adding two Lagrange multipliers ( the price of having an unconstrained formulation ) to the standard action for the p-branes. We shall follow an alternative approach by embedding the theory of strings( p-branes) into a larger space 
 that we call {\bf C}-space. We find that the principle of covariance in such a  {\bf C}-space forces upon us a drastic new view of spacetime and extended objects and, consequently, an extension/modification  of the ordinary QFT that we are acustomed to. In particular, modifications of the Heisenberg uncertainty principle \cite{Oda} , \cite{Li}, \cite{Cas1} and the notion of an observer-dependent Hilbert space \cite{Rov}

The Klein-Gordon-like Loop Master field $\Phi [X^{MN}]$ will depend now on an antisymmetric matrix ( to be defined shortly) living in an enlarged space that we shall call {\bf C}-space ( some sort of superspace not to be confused with the one in supersymmetry). The elements of this {\bf C}-space , a Category in the modern language, are comprised of a collection of $histories$ of point-particles, closed lines (loops), bubbles ( world-sheet surfaces), lumps, p-branes,  of arbitray topologies, $D$-branes.... The bubbles may have also a boundary $\gamma$ : i.e 
a disk topology with any number of handles attached to it. The lumps ( world volume of a membrane) may have two-dimensional boundaries and so forth and so forth.  
For simplicity and without loss of generality we shall study the simple case of points, loops and bubbles only.

The $interacting$ master field theory associated with $\Phi [X^{MN}]$ ( corresponding to a given quantum state $|\Phi>$ ) will involve :

(i) bubbles  of non-zero areal time $A$ (of arbitrary topologies) emerging from the vacuum ( zero orgin of coordinate-matrices  in this {\bf C}-space );i.e 
a virtual closed string history. These bubbles will also carry with them their center of mass coordinates.

(ii) An open surface with boundary $\gamma$ and proper areal time $A$, with boundary holographic coordinates $\sigma^{\mu\nu}(\gamma)$ ( shadows onto the spacetime coordinate planes ); i.e it will involve disks of areal time $A$,  with boundary $\gamma$
with any number of handles attached to it  emerging from the vacuum. Thse open surfaces will also carry with them their center of mass coordinates ( inside). 

(iii) loops emerging out of the vacuum , with shape $\gamma$, with a given center of mass $x^\mu_{CM}$ in spacetime; i.e a sort of `''loop-instanton''  in $areal$ time $A$.

(iv) null world sheets of areal time $A=0$ ; i.e null/tensionless closed string-histories  with null boundary shape $\gamma_{null}$. These null world sheets will also carry with them their  center of mass coordinates : a null line in spacetime.

(v). It will, in addition,  involve point particle-world lines ; i.e the zero modes or  center of mass coordinates of the loops and open/closed surfaces. 

(vi) the master field will in general create an ``object'' that encompasses all of the above
for the most general value of the matrix $X^{MN}$ ( defined shortly). 

In the case of a membrane one requires a rank three antisymmetric tensor : $X^{MNP}$ and for a $p$-branes a rank $p+1$ antisymmetric tensor. 
All these histories are observer dependent in {\bf C}-space. 
For example, an observer moving very fast
( ``infinite-momentum `` frame)  in this {\bf C}-space will see the original loops 
and bubbles 
(observed by the first static observer in  {\bf C}-space ),  
''contract'''  to a point 
and to a line, respectively. 
The former single point must not be confused with  the zero-matrix origin of coordinates, the vacuum. Therefore, ``Special Relativity `` in this {\bf C}-space will transform the above objects into one-another and combinations thereof. In this  fashion the $categorical$ aspects of such space will be manifested ; i.e topology change in particular.

There is an infinite number of reference frames linked to each observer. As each single one of  
these infinite  observers  explores, from his/her own reference frame, each event in {\bf C}-space labeled by a ``point'',  an antisymmetric matrix $X^{MN}$, he/she will see many histories. Depending on the values of $X^{MN}$ : he/she will see a point-history;  a loop-history or open 
surface with boundary $\gamma$ ;   a bubble ( a virtual string history);  a virtual loop 
( a `'loop-instanton'' in areal time $A$ which is $not$ the same as the instanton in coordinate imaginary time ) ....  
Another observer will see a different scenario from a different frame of reference.

{\bf C}-space is the arena for all these histories or events in the language of John Wheeler. An observer in {\bf C}-space witnesses the ``history of all possible histories'' associated with points, loops, bubbles, depending on where he/she is ``sitting''. 
It is $in$ this {\bf C}-space,  a category comprised of point, loops and bubbles,    where one quantizes spacetime . 

Recent recents based on the Loop Quantum Gravity formalism,  have found that  
the length, area, volume operators have discrete eigenvalues 
\cite{Rov}. This seems to suggest that 
the true quanta of spacetime are now strings of Planck length, loops of Planck-area, bubbles of Planck volume, etc.....This also appears to be consitent with the ideas that there seems 
to be an  $alleged$ minimum distance, minimum area, minimum volume,...in Nature implied by the stringy-uncertainty principle . There is no such a thing as a point in this {\bf C}-space. The only dimensionless ``point'' $is$ the vacuum ( zero matrix) which is the zero event ( the origin)  from which all histories emerge/spring from. It must not be confused with the center of mass coordinates of the loops, bubbles!

 Evidence accumulates today which seems to suggest that that 
 spacetime is $fuzzy$ at 
Planck  scales : one cannot resolve a point with an  infinite accuracy. This is what the modified stringy-uncertainty principle and Noncommutative geometry 
have taught us  \cite{Lan}, \cite{Cas1},  
\cite{Oda}, \cite{Li}. 
There is an alleged  minimum resolution-distance in Nature \cite{Not}, and consequently, a new principle of Relativity : The theory of Scale Relativity proposed by Nottale where the Planck scale is the minimum  $resolution$ attained  in Nature. These spacetime resolutions $are~not~statistical$ uncertainties of the 
spacetime coordinates.  They are the natural $extra~resolutions$ one has to introduce in a truly Fractal Space Time. The latter space requires both coordinates $and$ resolutions. It is roughly 
speaking like a sort of generalized phase space $(q,p)$ of $double$ dimensionality than the configuration, $X^\mu$ spacetime.   
Increasing evidence seems to point that the Universe appears  to be 
fractal at very small and very large scales.  

We must $not$ confuse the string length scale $l_s$ with the Planck length, 
$l_p$ and $resolutions$ of spacetime intervals, that a physical apparatus can resolve,  with  with the actual value of the spacetime interval.  For example, a closed loop of radius $l_p$ will have for Euclidean perimeter a length $l_s = 2\pi l_p$ and area $  \pi l_s^2 >\pi l_p^2 $. $D$-branes, can probe distances smaller than $l_s$ \cite{She}. Scale Relativity suggests that it takes an $infinite$ amount of energy to attain Planck scale $resolutions$. Like ordinary motion Special Relativity sets a maximum speed in Nature, the speed of light, in Scale Relativity there is a minimum resolution 
distance. If a massive object were to attain such speed of light it would consume an $infinite$  amount of energy. A hypothetical infinite Lorentz contraction ( null like ``observer'') of a line segment to zero does $not$ imply that the $resolution$ at which one observes the length of zero is also equal to zero! For example, the value of $q=0$ in phase space does not imply that the value of $p=0$. Scale Relativity in a fractal spacetime imposes  upon us the need to enlarge our 
ordinary concept of a smooth spacetime to include extra independent degrees of freedom : the resolutions.  

Ordinary Geometrical concepts do not apply at those (fractal) scales where the Hausdorff dimensions can attain an $infinite$ value. A phase transition occurs where one has pure topology : there is no notion of distance, below the Planck scale : $<g_{\mu\nu}>=0$.  It is the result of the breakdown of this topological symmetric phase \cite{Oda}, \cite{Li} , at the Planck scale and beyond, when the notion of metric emerges : $<g_{\mu\nu}>\not=0$
 and, consequently, a distance, light-cone, particle propagation, dynamics,...is possible etc....    

Furthermore, what is more intriguing, if one chooses $D=4$ as the spacetime dimension where all these events take place, and one considers $trivial$ topologies corresponding to stringy-tree level amplitudes ( spherical bubbles, disks with no handles,...) , we can show that the dimension of this {\bf C}-space associated with the space of antisymmetric matrices, $X^{MN}$,  will be precisely $11$ dimensional . We have at this stringy-tree level point a  
$D=4$ spacetime unified description of points, closed lines (loops)  and closed/open surfaces of trivial topology  ,  in an $11$ dimensional {\bf C} space.

We will write down one Master Covariant Loop Wave Functional Equation describing $all$ the quantum dynamics of both points, closed lines and closed/open surfaces $combined$.   
We will also have a holographic description of the quantum dynamics in this extended loop space ( of points, loops and bubbles ) given by their projections/shadows (Plucker-coordinates) from the {\bf C}-space ( essentially an extended loop space) onto the $D=4$  spacetime. We have a {\bf C}-space whose {\bf C}-events ( point, loop, bubble histories) are labeled by Noncommutative antisymmetric matrices. And finally, $smooth$ topology change appears  naturally as one moves from ``point to point'' in this {\bf C}-space.  

 After this introduction we proceed to write down a Covariant Loop Wave Functional Master equation for the master field $\Phi$. Firstly, 
one needs to coordinatize the areal time, the $\sigma^{\mu\nu}(\gamma)$ and their  center of mass coordinates into one single object ( the analog of a four vector). Therefore,  
we incorporate the $A$, $\sigma^{\mu\nu}$ and center of mass coordinates coordinates into one 
$(2D)\times (2D)$ antisymmetric matrix whose block components are :

$$X^{MN}\equiv \{  X^{ab}=- X^{ba}.~a,b=0,2,3,...D-1.~~ X^{\mu\nu}=-X^{\nu\mu}; 
~~ X^{a\mu}=-X^{\mu a}\}. \eqno (5-1 )$$
writing explicitlty this matrix :

$${\bf X}^{MN}=\left (\begin {array} {cc} X^{ab} &  X^{a\mu}\\
X^{\mu a} & \sigma^{\mu\nu} \end{array} \right )\ \eqno (5-2 )$$

The spatial loop holographic components are the usual $X^{\mu\nu}\equiv \sigma^{\mu\nu}(\gamma)$. They make up a matrix lying on the south east block corner.  The  
$D\times D$ matrix $X^{ab}=- X^{ba}$ , lying on the north-west block corner of $X^{MN}$, has  for the 
two nonzero components in the two  off-diagonal corners  the values of $A, -A$ respectively. 
$A$ in the upper right-hand corner and $-A$ in the lower left-hand corner, respectively. 
Finally, the two matrices $X^{a\mu}=-X^{\mu a}$,  north-east and south-west blocks,  are the remaining ones required to fit the center of mass coordinates $x_0,x_1,.....x_{D-1}$. We do this by 
assigning  the (non-zero)  diagonal components of $X^{a\mu}$ the values of $x_0,x_1,.....x_{D-1}$ and for 
$ X^{\mu a}$ the values of $-x_0,-x_1,-x_2,....., -x_{D-1}$ to assure that the $X^{MN}$ is indeed antisymmetric.

We notice that the elements of the matrix have dimensions of area and length. It proves convenient to rescale all the elements of the matrix in such a way to have a matrix of $area$ dimensions. One should then leave fixed the areal time  $A$  
and the holographic shadows, $\sigma^{\mu\nu}$ and rescale the center of mass coordinates by a factor of $l_P$ ( Planck's length).  This will be important when we construct  a metric in {\bf C}-space that will define an interval of dimensions of $area$. It will also be very important to establish the  correspondence principle. We will see that in the limit of $l_P\rightarrow 0$ everything collapses to a ``point''( leaving only the commuting matrices $X^{a\mu}$ containing the center of mass coordinates on their diagonal )  and we recover the ordinary Minkowski spacetime interval from this limiting process of the interval in {\bf C}-space.

The antisymmetric matrix $X^{MN}$ encodes the histories of points ( center of mass) , loops and bubbles ( with/without boundary)  in one single footing. Also  
the formation  of $null$ loops of zero proper areal time, $A$  ; i.e a $tensionless ~string$ appear here as well.   Tensionless strings must be present in this covariant formalism . 

For example, In $D=4$ for trivial topologies ( tree level stringy amplitudes) , we have an $8\times 8$ antisymmetric matrix. There are ${4.3\over 2}=6$ independent  components for the holographic shadows of the loop. There are four components for the center of mass 
coordinates and one component for the areal time $A$. The total number  of independent matrices is $11$. Any $8\times 8$  antisymmetric    $X^{MN}$ can be expanded into a basis of $11$ $8\times 8$ 
antisymmetric 
matrices,  $E^{MN}_i$ as  $\sum a_iE^{MN}_i.~i=1,2,3...11$.      

One may be inclined  to add extra matrices, $X^{MN}_h$, to the elements of the {\bf C}-space,  to account for the self-interactions,  like stringy-loop interactions,  by including surfaces of arbitrary number of handles all the way to infinity  and , in general, nodal singularities ( pinching a handle to a point). An augmented moduli space. 
 
However, interactions in string theory are mediated by topology. 
The four string tree level $S$ matrix amplitude can be computed as the vacuum correlation function of four Vertex  operators inserted at four punctures :  
on-shell emission or absorption of particles by the interpolating world sheet having as boundaries the locations of the strings. The latter can be conformally mapped to a world sheet with four punctures.   
Stringy-loop perturbative corrections are obtained by adding an arbitrary number of handles to the world sheet. In the string field theory language, for example,  the 
three string vertex interaction requires to use $\Psi^* \Psi^*\Psi $ string-field interactions. 

Since the $interacting$ field theory of the quantum master field 
${\hat \Phi}  [X^{MN}]$ should   
already encode the introduction of $all$ sort of stringy-tree level interactions and stringy-loop corrections , it won't be necessary to add extra degrees of freedom to the {\bf C}-space by using a set of $X^{MN}_h$, for $h=0,1,2,....\infty$. The interacting quantum master field theory will take into account all of that from the start. It will create an arbitary number of handles, boundaries, holes,.... 

For example; it will involve  two bubbles interacting at one point and then merging into a third bubble or vice versa :  the breaking of one big bubble into two smaller bubbles, 
at one point. These  ``elastic''  bubbles
of Planck volume will be our quanta of volume in spacetime ; the ``elastic'' closed loops of Planck area will be our areal-quanta of spacetime ,....two fundamental bubble-units ( two separate volume quanta) can merge into one bigger bubble of twice the Planck volume ; $n$ bubble units ( $n$ quanta) can merge into one bigger bubble of $n$ times the Planck volume ...and vice versa :
one big bubble of $n$ times the Planck volume can break into $n$ separate smaller bubbles of one unit of Planck volume. Similar arguments apply to the quanta of Planck area : 
two closed loops interact at one point and merge into a third loop of twice the area. Or vice versa : the big loop of two quanta of area breaks at one point into two smaller loops of one quanta of area.

As in ordinary quantum field theories , the master field propagator is obtained from the kinetic terms in the action and the form of vertices are determined through the types of interactions one constructs for 
the master field.  
 In the true $categorical~sense$,  the $interacting$ quantum master field theory will play the role of a $categorical~functor$ that will map two incoming closed strings ( of one areal quanta) into a third string ( of two areal quanta) ; a torus ( of one volume quanta) into a sphere ( of one volume quanta);  a two-handled Riemann surface ( two units of volume quanta) into two  
torus ( of one unit of volume quanta) , etc...
It is in this way how  $topology ~change$ will be captured , in a $smooth$ fashion from the $categorical$ point of view only , by the interacting quantum dynamics of the master quantum field  ${\hat \Phi}  [X^{MN}]$    $in$  the 
{\bf C}-space. However, the topology change it is $not$ smooth from the ordinary spacetime point of view !. The torus-sphere transition by pinching the  handle of a torus to a nodal point is not a smooth transition in spacetime. We must emphasize the  the Master QFT   
we are constructing is $in$ {\bf C}-space and $not$ in spacetime !

Segal \cite{Seg} has advocated this  Categorical picture to 
reformulate the physicist's notion of $2D$ Conformal Field Theory in the Mathematical language of functors and categories. For example, the three-string vertex interaction admits a very natural functorial interpretation as the functor that maps two incoming loops to a third outgoing loop. The functorial  map plays exactly the same role as the $interpolating$ string world sheet ( of arbitray topology) having for boundaries the three loops with appropriate orientations. The quantum Master field 
${\hat \Phi} $ will play an analogous role as Segal's functorial map. It will generate surfaces of arbitary interpolating topologies among any number of incoming/outgoing strings ( the boundaries).  
Crane \cite{Cra} has proposed also to view 
quantum gravity as categorical ( algebraic) process. Smolin has developed   
Penrose's  Spin-Networks program \cite{Smo1} as an algebraic approach to quantum gravity and also a more recent holographic formulation of quantum general relativity in terms  of states and operators defined on finite boundaries
\cite{Smo2}. For another interesting approach one has the work of \cite{Bae2} based on Spin-Foam models and the Discrete and Combinatorial Geometric approach of Matroids \cite{Nie}.

Open strings and $D$-branes  will be incorporated later . Timelike Wilson loops represent quark-antiquark pair creation and annihilation . These quarks/antiquarks are attached to the end points of two open strings. One moving forward in time and the other backwards in time. 
Hence, time-like Wilson lines correspond to mapping the boundaries of an open string world-sheet into the quark-antiquark worldlines. 
Null-like loops represent tensionless strings and spatial loops the usual ones. These were mentioned above.

The $zero$ origin of matrix-coordinates represents the $vacuum$; i.e the $zero$ matrix. It can be ``translated'' to another point in {\bf C}-space by the action of a ``constant-vector  translation'' as it occurs with ordinary Minkowski spacetime under the action of the 
Poincare translation operators.    
It is with respect to this zero origin that we label our coordinates, $X^{MN}$.

For example, 
a  spatial loop can be created out of the vacuum ( zero matrix)  and, in a given areal time, $A$, can appear at a given region, $\gamma$,  of the usual target spacetime onto which we embed the loop. 
This history  emerging from the vacuum is represented by the matrix $X^{MN}$ ( relative to the zero matrix). Another picture is that  the spatial loop 
could shrink to zero size at the end of the road :  we will have a closed surface or bubble of areal time $A$ 
( a virtual closed string) .  
Another possibility is to have the  spatial loop shrink to zero from the very beginning and what we have is the world-line of a point particle ( center-of mass motion) from the origin to another point; etc...

If we view the origin ( the vacuum)  of coordinates attached to one particular observer ( like any 
spacetime coordinate assigned to any spacetime event ) as the zero matrix, under the analog of a Lorentz relativistic transformation  it will be transformed  also into the zero matrix. An additional   ``Poincare translation'' will shift the latter zero matrix to a 
non-zero ( constant) matrix that we may  label as 
the ``transformed vacuum'' under the full Poincare-group-analog  ( Lorentz plus translations).  
Therefore, the notion of the so-called vacuum ( origin of coordinates)  is $observer$ dependent for this  general case . When one speaks of a ``loop'' as an element of a Loop space,  one must talk of loops $based$ at a fixed $point$. In a similar fashion, here one must specify , first, the point/origin to which one assigns the beginning/birth of the  history of a 
bubble, loop, point.... 

For example,  viewing a virtual closed string as the  spatial loop history  emerging from the vacuum ( origin of coordinates) and ending at another point ( shrinking the closed string to zero) in a given areal time $A$,   as a bubble ( baby universe) history, from the point of view of two different observers in {\bf C}-space,  one has two different pictures :   

The action of the Poincare-group-analog, will transform one  bubble (universe) history as seen by the first observer, into a $new$ bubble (universe) history as seen by the second observer . And the latter history , can be seen as the one which has been  created out of a ``translated vacuum'' . Translated with respect to the old vacuum 
( from which the first bubble-universe history  emerged from). 
If the second observer is moving `''very fast'' with respect to the first  observer, he/she will 
see a $different$ virtual string history : a $different$  bubble that appears now  ``contracted''. For him/her now the bubble looks now more like a $line$ ( virtual point particle) instead of a closed surface (virtual closed string). We believe that this could be related to the plausible origins of Duality in 
String theory and one of the underlying principles behind $M$ theory : 

$Covariance$ in {\bf C}-space. It is the generalized principle of covariance in this categorical {\bf C}-space comprised of point histories, loop histories, bubble histories,.... 
it is this generalized principle of covariance ( area-preserving diffs in {\bf C}-space ) apply to 
the categorical {\bf C}-space , comprised of a whole collection of extended-loop spaces :  
points, loops, bubbles, lumps....,  that we believe may hold important clues in the underlying principle behind 
$''M~F~S~...''$ theory. It is relativity to its fullest potential : the relativity of histories.  
Each observer sees a different history/herstory accordingly to his/her motion in {\bf C}-space.  There is no preferred referential  frame of reference. All are equivalent. To one observer one history/herstory looks like that of a bubble; to another it looks like the one of a line;  to another like the one of a point.....all p-branes can be encompassed in this fashion into one single footing in one categorical space {C}-space. In the case that we have only points, loops, bubbles, 
their holographic shadows of an $11$-dimensional {\bf C}-world cast upon the ordinary $4D$ spacetime is what we observe. Plato's old view of reality.   

Summarizing, there are an infinite number of histories associated with one, and only one loop, $\gamma$,  since each  history may be viewed by an infinite number of observers.  Each history carries its own vacuum ( attached to each observer). The family of vacua are related by ``translations'' in {\bf C}-space. There is $no$ preferred universal inertial frame of reference in {\bf C}-space, and for this reason, there is $no$ universal vacuum.  In loop space there is $no$ prefered base point.    

We are ready now to implement this generalized relativity principle to the set of point, loop and bubble histories. The analog of an invariant proper time interval -in a fixed reference frame with a given zero orgin of matrix coordinates -between two matrices or histories emerging from the vacuum,  
$ X_{MN}(\tau);X^{LP}(\tau)$ ( ``points''),  whose evolution is parametrized by $\tau$   is :

$$(d\tau)^2 =      [dX_{MN}(\tau)]~{\cal G}^{MN}_{LP}[X^{AB}(\tau)]~ [dX^{LP}(\tau)]. \eqno (5-3 )$$
where ${\cal G}^{MN}_{LP}[X^{AB}(\tau)]$ is a {\bf C}-Metric  in the space of matrices 
$X^{MN}$. For the time being let us assume the {\bf C}-space is ``Flat''. The introduction of 
``parallel transport'' , ``curvature'', and the ``principle of equivalence'' will be the subject of another work.  

As mentioned earlier the {\bf C}-metric interval $\tau$ has $area$ dimensions. Let us now rescale the 
interval $(d\tau)^2$ by a factor of $l_P^{-2}$ leaving us with a quantity of dimensions of 
$(length)^2$, which is the dimensions of the standard spacetime interval in Minkowski spacetime. Upon absorbing the factor of  $l_P^{-2}$ into the $dX_{MN} dX^{LP} $ matrices as 
$ [l_P^{-1}  dX_{MN}] [l_P^{-1}dX^{LP}] $ one can rescale each one of our original matrices ( of $area$ dimensions) by a factor of $ l_P^{-1}$. The correspondence limit is defined as  
(i)$l_P\rightarrow 0$ and (ii) 
$A\rightarrow 0$ and $\sigma^{\mu\nu} \rightarrow 0$ . One should end 
then with a physics involving points only ; the bubbles and loops shrink to a point : the center of mass coodinates.    
We see that in this  correspondence limit,  the rescaled matrices  $l_P^{-1}  X_{MN}$ tend to a matrix whose only components will be those of the $X^{a\mu}$ containing the center of mass coordinates along their diagonals.  This is fairly clear because in the limit :
  $l_P^{-1} A= 0   $; $  l_P^{-1}\sigma^{\mu\nu}=0$ ( Number of Planck cells is kept finite) and    $  l_P^{-1}.(l_P.x_{CM})=x_{CM}$.     The latter limit is what one expects since we had rescaled earlier the $x_{CM}$ by a factor of $l_P$ so that all elements of our matrices have the same area-dimensions . The quanta of areas come in discrete multiples of the Planck area. \cite{Rov}. 

Therefore, in the above correspondence limit     (i)$l_P\rightarrow 0$ ; (ii)  $A\rightarrow 0$ and $\sigma^{\mu\nu} \rightarrow 0$ ( the Number of Planck cells is finite)  one ends up with a standard spacetime interval of 
dimensions of $ (length)^2$ in Minkowski space involving  point-histories  only. The anticommuting matrices ``collapse'' to pure diagonal $commuting$ ones. Taking the  $l_P\rightarrow 0$ is tantamount 
of going from noncommutative ( fuzzy) geometry to a commutative one ( well defined points). 
It is the analog  of taking the zero slope $\alpha' =0$ limit in string theory, the field theory limit. Quantum effects in perturbative string theory are seen as a topological expansion in the genus of the Riemann surface. Since we wish to capture the full Nonperturbative physics of string theory we must follow a different approach. At the end of this  section we will propose the Moyal-Fedosov deformation quantization 
technique to construct the putative Nonperturbative interacting quantum field theory for the master field in {\bf C}-space. This is still very preliminary.

One plausible outcome of this formalism is that the fundamental orgins of gravity from string theory would no longer appear  to be such a mystery : Gravity in spacetime emerges from the string as a result of the geometrization of the ``parent'' {\bf C}-space onto which we embed the string theory itself. Furthermore, the initial spacetime with coordinates $X^\mu$, could in principle be any background spacetime. So this formulation will also be truly background independent.  The simple fact that the spin two graviton appears in the perturbative string spectrum about a flat spacetime background is $no$ satisfactory explanation of $why$  gravity emerges from string theory. The background independent formulation is still missing. We hope that this approach based on loop spaces may shed some light into this problem.    

$\tau$ is a parameter or proper-time  clock that is $intrinsic$ to the {\bf C}-space manifold of antisymmetric matrices ( evolution of the events that keep track of the history of histories emerging from the vacuum) . There are now many ``times''\cite{Vaf} , 
\cite{Bars}. The invariant proper time $\tau$ ( intrinsic clok of the {\bf C}-space) defined above in eq-(3-26).    
The string world sheet $areal$  time $A$. The world sheet $coordinate$ time $\sigma^0$. And the target spacetime coordinate $X^0$.  

The analog of a ``Lorentz'' invariant interval  is :

$$  (d\tau')^2=(d\tau)^2 =     dX_{MN}(\tau)~{\cal G}^{MN}_{LP}[X^{AB}(\tau)]~dX^{LP}(\tau)=$$
$$    dX'_{M'N'}(\tau')~{\cal G}'^{M'N'}_{L'P'}[X'^{A'B'}(\tau')]~dX'^{L'P'}(\tau'). 
\eqno (5-4 )$$
and the corresponding finite ``Lorentz''  transformations implemented by the matrices 
$\Lambda^{MN}_{M'N'}$ is :

$$X'_{M'N'}=   X_{MN}\Lambda^{MN}_{M'N'}(\beta_i)         .~~~
X'^{L'P'}=(\Lambda)^{-1~L'P'}_{LP} (\beta_i)X^{LP}.~   \eqno (5-5 )$$
where $\beta_i$ are the ``boosts, angles..'' parameters associated with each one of the generators of transformation group. As said above, the ``Poincare-translations'' will shift the vacua ( origin of matrix-coordinates). 

The invariant metric, ${\cal G}={\cal G}'$ must  transform as  :

$${\cal G}'^{M'N'}_{L'P'}= (\Lambda)^{-1~M'N'}_{MN}(\beta_i)   {\cal G}^{MN}_{LP} 
\Lambda^{LP}_{L'P'}(\beta_i)=  {\cal G}^{M'N'}_{L'P'}. \eqno (5-6 )$$
Notice that the proper-time interval in {\bf C}-space has units of $area$. This will be important as discussed shortly.  If one wishes to include membranes, three-branes,...p-branes into the 
{\bf C}-space, one will need antisymmetric objects of any $p+1$ rank : $X^{MN.....P}$. As we have said earlier,  a new formulation of all (bosonic) p-branes as Composite Antisymmetric Tensor Field Theories of the volume-preserving diffs group, was given in  \cite{ Cas4}. The advantage of this formulation is that one has $S$ and $T$ duality already built in from the very beginning.  There is no need to conjecture it.   

The metric must be antisymmetric under the exchange of $M,N$ and under the exchange of $L,P$ and also symmetric under an exchange of the two pairs $(MN)\leftrightarrow (LP)$. 
The symmetry group that leaves invariant the ``distance'' between two antisymmetric matrices (
``area'' between two histories) is the 
Lorentz-group-analog that characterizes the notion of relativistic invariance in the space of 
$X^{MN}$. This {\bf C}-space is  endowed with a Lorentz-invariant-analog of the  metric ${\cal G}^{MN}_{LP}$. 

The main question now is : What is this symmetry acting in {\bf C}-space that replaces ordinary spacetime Lorentz invariance ? It is the symmetry acting on the {\bf C}-space which is the generalization of the area-preserving diffeomorphisms group.  However these ``Areas''  are areas in {\bf C}-space not in spacetime. It is more closely related to a topological symmetry instead of a spacetime symmetry \cite{Oda} , \cite{Li}. 

The action for these matrices in {\bf C}-space should be the analog for the point particle :

$$  S= {\cal M}^2\tau= {\cal M}^2 
\int  \sqrt {  dX_{MN}(\tau){\cal G}^{MN}_{LP}[X^{AB}(\tau)]~dX^{LP}(\tau)}. \eqno (5-7 )$$
The analog of invariant proper mass is now a tension-like quantity , ${\cal M}^2$. The massless case corresponds to tensionless strings ( null cone analog). $\tau>0; \tau<0$ will be timelike or spacelike analog respectively. 
The temporal evolution wave equation is then :

$$i {\partial \Psi [X^{MN}, \tau] \over \partial \tau} =
{1\over {\cal M}^2} {\delta^2   \Psi [X^{MN}, \tau]\over \delta X^{MN}  \delta X_{MN}}. 
\eqno (5-8)$$

Setting as usual : 

$$   \Psi [X^{MN}, \tau]=\Phi [X^{MN}]e^{-iS}= \Phi [X^{MN}]e^{-i{\cal M}^2 \tau}. \eqno (5-9)$$
and inserting it into the temporal evolution equation yields the Klein-Gordon-like Covariant  Loop Wave Master equation :

$$ {\delta^2 \Phi [X^{MN}]\over \delta X^{MN}  \delta X_{MN}}-   {\cal M}^4  \Phi [X^{MN}]=0.   \eqno (5-10  )$$         
where now the areal times, $A$ and spatial areas ( coordinates) $\sigma^{\mu\nu}(\gamma)$ are treated on the same footing. Extra variables $X^{\mu a} =-X^{a\mu}$ must also be added which represent the center of mass motion coordinates. Eq-(5.10) can be generalized to all 
$p$-branes by simply working with antisymmetric tensors of rank $p+1$ and inserting the $p$-brane tension squared for the ``mass'' squared  terms.

If one chooses a complex valued field, $\Phi [X^{MN}]$, one could reinterpret it as the 
field that creates loops, bubbles, point-particles and its complex conjugate,  the field that will create anti-loops , anti-bubbles, anti-particles . The masless master field case , ${\cal M} =0$ corresponds to the whole array of processes like  the  ceation/annhilation  of 
$null$ loops ( boundaries of the null world sheet of tensionless closed-strings), null bubbles , null lines for  massless point-particles....  
The Master Loop Wave Eqation equation can be obtained from a $free$ field theory  action 

$$S[\Phi]={1\over 2} \int [DX^{MN}] {\delta \Phi^{\dagger}  [X^{MN}]\over \delta X^{MN}} 
{\delta \Phi [X^{MN}]\over \delta X_{MN}}
+{\cal M}^4  \Phi^{\dagger} [X^{MN}] \Phi [X^{MN}]      .        \eqno ( 5-11 )$$     
Interactions  of the type $\Phi^3, \Phi^4...$ can be also be added, similar to those in Closed String Field Theory \cite {Bar}.  
These terms  will provide the interacting field theory of the master field. Self interactions naturally introduce in string field theory the creation of handles. This corroborates that the full interacting field  theory of the master field will be responsible for the creation of bubble-histories  of arbitrary nontrivial topologies with/without boundaries. As said above, it is for this reason that one does not need to add extra matrix coordinates to the {\bf C}-space. The interacting field theory will reproduce all these terms.   

To sum up : we are proposing a plausible Master  
scalar field theory in {\bf C}-space that upon quantization ( to be discussed shortly) describes  the Quantum Non-perturbative dynamics of 
points, loops and bubbles encoded into a single antisymmetric matrix, $X^{MN}$. 
In $D=4$ , for trivial topologies ( stringy-tree level) , this {\bf C}-space is a Noncommutative $11$-dimensional space of $8\times 8$ antisymmetric matrices ;  it implements the holographic principle ( proyection of the 
{\bf C}-shadows onto the spacetime coordinate planes) and its quantum dynamics is governed by a covariant loop wavefunctional equation of the Klein-Gordon type in the free field case. 
The interacting master field theory will automatically introduce objects of arbitrary topologies. And finally, the correspondence limit , $l_P \rightarrow 0,...$ reproduces the standard spacetime proper interval after rescaling the area interval by $l_p^{-2}$ , absorbing $l_p$ powers into the matrices $X^{MN}$   and taking the $l_p=0$ limit. 

``Square-roots'' of the latter free field master equation will reproduce, a la Dirac, a ``spinorial'' Master  wave equation. Doing the same square-root procedure for the ``non-relativistic'' Schroedinger Loop Wave equation will yield the analog of Supersymmetric Quantum Mechanics. This is no surprise : the M(atrix) models \cite{Ban}, infinite momentum frame of $M$-theory,  are essentially a  Supersymmetric Gauge Quantum Mechanical Model of the area-preserving diffs group $SU(\infty)$.

The fact that area-preserving diffs in this {\bf C}-space of $2D\times 2D$ antisymmetric matrices ( for trivial topologies) is the corresponding symmetry algebra ( the relativistic Lorentz analog)  , suggests that the $symplectic ~group$ ( and its supergroup version which contains the conformal group) must play a fundamental role. Symplectic geometry and Moyal-Fedosov Deformation Quantization were used in \cite{Cas2} to describe $W_\infty$ Geometry.  The space of $2D\times 2D$ antisymmetric matrices is a generalized symplectic space. The generalized area is associated with the generalized symplectic form. It seems that it all boils down to symplectic geometry in higher-dimensional loop spaces. The holographic shadows of these higher-dimensional loop spaces ( endowed with a symplectic structure)   are the point-histories, the  loop-histories, the bubble-histories,... the lump-histories, ...extended object-histories  that we see in spacetime.

How will one implement, if possible,  a Moyal-Fedosov Deformation Quantization program to the interacting quantum master field theory action  ? Assuming that one has an invertible non-degenerate symplectic structure which allows to compute ``Poisson brackets'' in loop spaces , that the notion of phase space is well defined, ...... The formal steps would be  :

(i) Given an operator-valued field ${\hat \Phi}[X^{MN}]$, living in the {\bf C}-space of 
$X^{MN}$ ,  and belonging to a Hilbert space of 
square integrable functions of $Q^{MN}$ , that we call ${\cal H}$, and whose phase space, 
${\Gamma}$ ,  is represented by the 
$Q^{MN}, P^{MN}$ ``coordinates'', the formal Weyl-Wigner-Moyal symbol map ${\cal W}$ takes :

(ii)    ${\cal W} : {\hat \Phi}[X^{MN}] \rightarrow { \Phi}[X^{MN}; Q^{PM},  P^{MN}  ]$.
It maps operator-valued objects of the Hilbert space,  ${\cal H}$,      
into smooth valued functions in phase space : the $symbols$ of operators. 

And :

(iii)     ${\cal W}( {\hat \Phi}[X^{MN}]{\hat \Phi}[X^{MN}]) 
\rightarrow { \Phi}[X^{MN}, Q^{MN}, P^{MN}  ] *{ \Phi}[X^{MN},   Q^{MN},    P^{MN}  ]$.
The symbol of the ``Weyl ordered''  product of two operators is mapped into the Moyal product of their symbols. The Moyal product is taken w.r.t the $Q,P$ phase space coordiantes.  

The Moyal Deformed Lagrangian density  of the interacting Master field action reads :

 $${\cal L} [\Phi (X, Q, P)]= 
{1\over 2} {\delta \Phi [X,Q,P]\over \delta X^{MN}}* {\delta \Phi [X ,Q, P]\over \delta X_{MN}}
+$$
$${ {\cal M}^4\over 2}   \Phi [X, Q, P] * \Phi [X,Q,P] + 
{1\over 3!} \Phi [X, Q, P ] * \Phi [X, Q, P]* 
\Phi [X, Q, P]+$$
$$ {1\over 4!} \Phi [X, Q, P ] * \Phi [X, Q, P]*\Phi [X, Q, P] *\Phi [X, Q, P]+........+{1\over n!} \Phi [X, Q, P] *....*_n \Phi [X, Q, P]. \eqno ( 5-12 )$$                        

This very formal Moyal deformation quantization of the interacting quantum master field theory 
using noncommutative Moyal star products w.r.t the variables $Q,P$, will provide , 
formally, the Moyal quantization  of the master field theory living in the noncommutative space of antisymmetric matrices $X^{MN}$.  
As far as we know, the problem of   $quantization$ of $noncommutative$ geometries has not been solved. This is a fundamental question that we hope could  be solved via a formalism similar to the one described by eq-(3.34). 

The action eq-(3.34) $resembles$ Witten's open string field theory action using the BRST quantization formalism with ghosts, antighosts,  \cite{Wit1} and to  Zwiebach \cite{Bar} closed string field theory action. The latter requires the use of the full machinary of the Batalin-Vilkovski antibracket formulation, the quantum master field equation, homotopy Lie algebras, operads,...........

These results (3.34) are very preliminary. All we wish to point out is that if, and only if,  deformation 
quantization techniques are indeed $valid$ in this context,  then they could be very useful in quantizing ( if possible) the interacting field theory of the Master field ${\hat \Phi} [X^{MN}]$ , in these {\bf C}-spaces. Such interacting Quantum Master Field theory 
will encode, in principle,  the quantum dynamics of a unified field theory encompassing gravity. It will provide a quantization of a master field living in a Noncommutative Geometrical space of antisymmetric matrices.

Early constructions of generalized spaces to describe non-perturbative string physics include the Universal Moduli space construction of Friedan and Shenker based on 
holomorphic vector bundles and flat connections constructed on the moduli space of infinite genus surfaces with nodes . Sato's Universal Grassmanian which has been central to the theory of integrable systems, where all Riemann surfaces with a puncture and a local coordinate system of all genera appear on the same footing. This is similar to our categorical construction with the difference that we include loop and point histories  also
into the picture. An approach closely related to the loop-space picture was analyzed by 
Bowick and Rajeev using the co-adjoint orbit method quantization of   
${Diff~S^1\over S^1}$. The latter is a Kahler manifold and the cancellation of the tangent bundle ( ghosts) curvature  against the vector bundle curvature (matter) was equivalent to finding a 
globally defined vacuum in the space  of orbits  : the matter plus ghost central charges vanishes  in $D=26$. Co-adjoint orbit methods can also be used for the area-preserving diffs orbits of 
$W_\infty$. See \cite{Bou}, \cite {Pho} for references. 

More results shall be presented elsewhere.

In short, the generalization of the principle of covariance-  
under area-preserving/volume-preserving diffs-       
applied to a categorical  
{\bf C}-space of extended-loops (points, loops, bubbles,.....histories ); i.e the construction of a generalized symplectic geometry in such categorical {\bf C}-space, 
seems to be, in principle,  a plausible appropriate ground to build $M$ theory.

 This has been $imposed$ upon us by simply covariantizing the ``non-relativistic'' 
Loop Wave Functional Equation ( Schroedinger) in order to obtain the Covariant Klein-Gordon-like Loop Wave equation for the single classical master field $\Phi [X^{MN}]$. Moyal-Fedosov deformation quantization techniques may be formally ( if possible) suited to quantize such classical master field theory in 
{\bf C}-space.   Spacetime Topology change, the Holographic principle are naturally incorporated in this formalism.
The fact that the Planck length $l_P$ is required to match units in the definition of the matrices $X^{MN}$ ( areal dimensions) and to establish the correspondence principle with ordinary Minkowski spacetime proper-time intervals of point events ( center of mass coordiantes) , in the
$l_P=0$ limit, suggests that   
the quantization of spacetime should come in fundamental discrete units of Planck length, area, volume,.... A $4D$ spacetime requires an $11$ dimensional {\bf C}-space of $8\times 8$ antisymmetric noncommuting matrices ( the only commuting matrices will be those containing the center of mass coordinates) for the special case that the categorical space contains solely points, loops and bubbles.

\vspace{5mm}

\vspace{0.5cm}
{\bf ACKNOWLEDGMENT}

The author wishes to thank Euro Spallucci and Stefano Ansoldi for stimulating discussions . We also  acknowledge enlightening conversations with Octavio Obregon in Guanajuato, Mexico at the early stages of this work  and to Devashis Banerjee, Miguel Cardenas for lengthy conversations  at the Abdus Salam ICTP in Trieste, Italy. We express our gratitude to  the Dipt. di Fisica Teorica of the 
University of Trieste for their  hospitality and support.

\end{document}